\pdfoutput=1

\documentclass[twocolumn,aps,prl,showpacs,superscriptaddress]{revtex4}

\usepackage{graphicx}
\usepackage{epsfig}
\usepackage{amsmath,amssymb}
\usepackage{color}
\usepackage{multirow}
\usepackage[colorlinks=true,citecolor=blue,urlcolor=blue]{hyperref}

\newcommand{\ket}[1]{|#1\rangle}
\newcommand{\bra}[1]{\langle#1|}

\begin{document}

\title{Device-Independent Certification of High-Dimensional Quantum Systems}



\author{Vincenzo D'Ambrosio}
 \affiliation{Dipartimento di Fisica, ``Sapienza''
 Universit\`{a} di Roma, I-00185 Roma, Italy}
\author{Fabrizio Bisesto}
 \affiliation{Dipartimento di Fisica, ``Sapienza''
 Universit\`{a} di Roma, I-00185 Roma, Italy}
\author{Fabio Sciarrino}
 \affiliation{Dipartimento di Fisica, ``Sapienza''
 Universit\`{a} di Roma, I-00185 Roma, Italy}
 \affiliation{Istituto Nazionale di Ottica (INO-CNR),
 Largo Enrico Fermi 6, I-50125 Firenze, Italy}
\author{Johanna~F.~Barra}
\affiliation{Center for Optics and Photonics, MSI-Nucleus on Advanced Optics, Departamento de F\'{\i}sica, Universidad de Concepci\'{o}n, 160-C Concepci\'{o}n, Chile}
\author{Gustavo~Lima}
\affiliation{Center for Optics and Photonics, MSI-Nucleus on Advanced Optics, Departamento de F\'{\i}sica, Universidad de Concepci\'{o}n, 160-C Concepci\'{o}n, Chile}
\author{Ad\'an Cabello}
 \affiliation{Departamento de F\'{\i}sica Aplicada II, Universidad de
 Sevilla, E-41012 Sevilla, Spain}


\begin{abstract}
An important problem in quantum information processing is the certification of the dimension of quantum systems without making assumptions about the devices used to prepare and measure them, that is, in a device-independent manner. A crucial question is whether such certification is experimentally feasible for high-dimensional quantum systems. Here we experimentally witness in a device-independent manner the generation of six-dimensional quantum systems encoded in the orbital angular momentum of single photons and show that the same method can be scaled, at least, up to dimension 13.
\end{abstract}




\maketitle


{\em Introduction.---}Dimensionality is a fundamental property of physical systems and a key resource in quantum information processing. Phenomena such as contextuality require systems of a certain minimum dimension to occur \cite{KS67,GBCKL13}; applications such as quantum secure communication have different levels of security depending on the dimension of the systems \cite{CBKG02,PW11}, and methods to characterize quantum states strongly depend on the assumed dimension of the systems \cite{WF89}. It is therefore of crucial importance to develop methods to certify whether a source produces systems that have {\em at least} a certain dimension and to distinguish quantum systems from classical systems of the same dimension. The first theoretical tools for providing lower bounds on the dimension of quantum systems were based on Bell inequalities \cite{BPAGMS08,VP08} and random access codes \cite{WCD08}.

Nevertheless, in cryptographic scenarios in which the preparation and measurements devices are not trustable, and also in scenarios in which the devices are imperfect or are not well characterized, it is desirable to asses the dimension of the physical systems in a ``device-independent'' (DI) manner \cite{ABGMPS07,GBHA10,BGLP11,RYZS12,BRLG13,BNV13,HYN13,HBS13}; that is, using only the correlations between preparations and the outcomes of different measurements and without making assumptions about the nature of the systems under observation or about the devices used to prepare and measure them. Theoretical tools achieving this goal where introduced by Gallego {\em et al.} \cite{GBHA10} under the name of DI dimension witnesses (DWs). DI DWs for systems of arbitrary dimension have been recently proposed \cite{BNV13}.

So far, DI DWs have only been used to experimentally certify the generation of classical and quantum systems of dimension 2 and 3 \cite{Nat12a,Nat12b}. However, realistic quantum information processing applications demand quantum systems of much higher dimensions. Moreover, the increasing technical complexity required for the generation of high-dimensional quantum systems turns DI DWs into fundamental testing tools.

The question is whether DI DWs are experimentally feasible to witness higher quantum dimensions or, on the contrary, the complexity of DI DWs may prevent their actual realization. For example, it can be shown (see below) that the number of parameters that have to be experimentally controlled to assess dimension $d$ in a DI manner increases as $2 d^2$. This reveals that assessing high dimensions in a DI manner constitutes an experimental challenge, since it requires much higher control and accuracy than previous experiments assessing dimensions in a DI manner \cite{Nat12a,Nat12b}. This is the challenge that we address here.

The aim of this Letter is to experimentally witness the generation of six-dimensional quantum systems and show that the observed results cannot be simulated with classical systems of dimension 6, using for that theoretical and experimental techniques which, as we show, can be scaled up to higher dimensions.


\begin{figure*}[tb]
\centering
\includegraphics[scale=0.48]{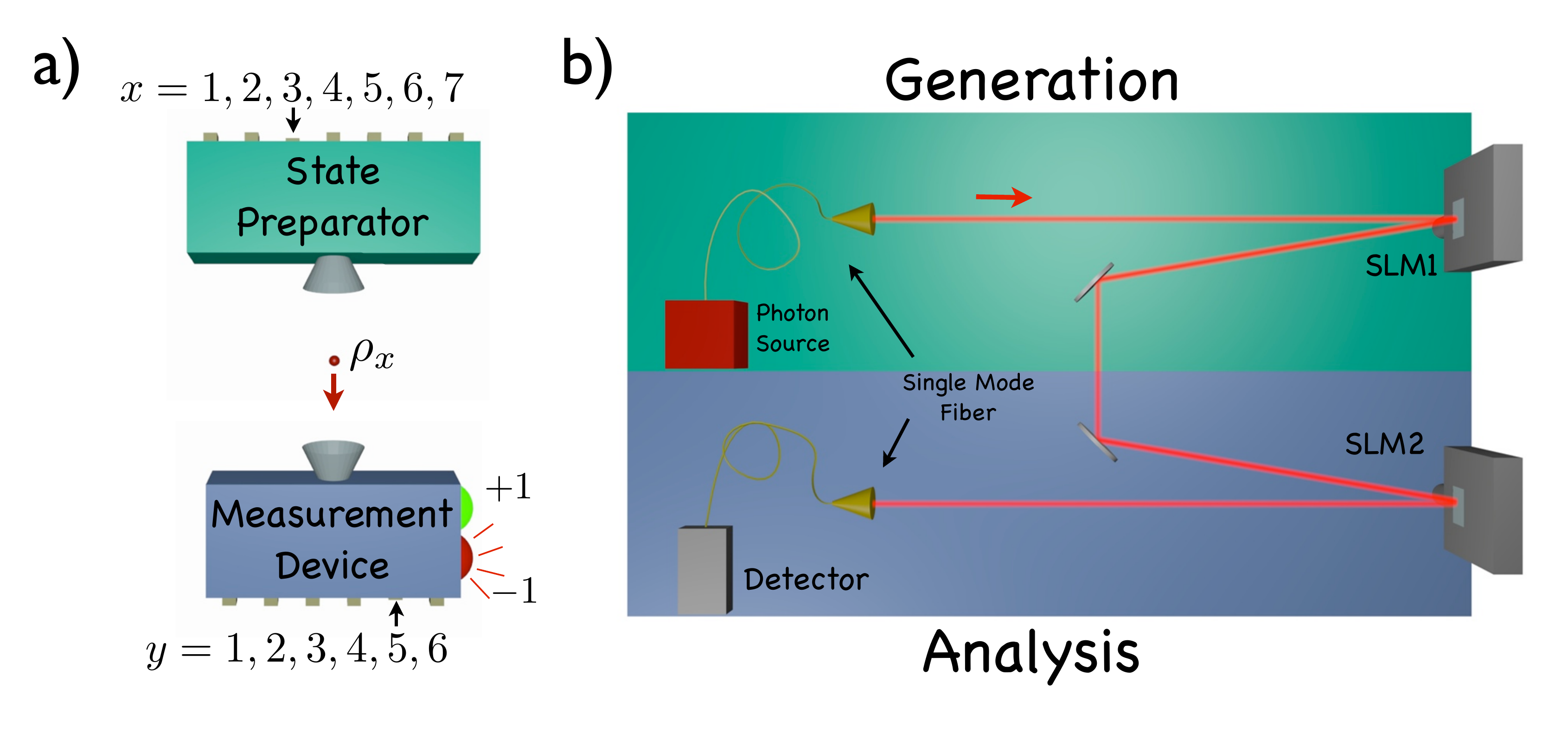}
\caption{\label{setup} (a) Schematic scenario for DI dimension witnessing: A physical system is prepared in a state $\rho_{x}$ chosen among a set of states $x$, and is sent to a measurement device. There, a dichotomic measurement, chosen among a set of observables $y$, is performed and the outcome is recorded. Specifically, the DI DW tested in our experiment requires 7 states and 6 observables. (b) Experimental implementation of the previous scheme: The setup consists of two stages, labeled as ``generation'' and ``analysis.'' In the generation stage, heralded single photons are produced by spontaneous parametric down-conversion in beta barium borate nonlinear crystal. The heralding photon is directly sent to a detector which acts as a trigger (not shown in figure), the signal photon is projected on the fundamental TEM00 Gaussian state ($l=0$) by means of a single mode fiber (Photon Source). The orbital angular momentum (OAM) state of signal photons is then manipulated with the spatial light modulator SLM1 in order to prepare each one of the 7 states required. In the analysis stage, projective measurements are performed by means of the spatial light modulator SLM2 in combination with a single mode fiber and a single photon detector. To avoid the Gouy phase shift effect, an imaging system (not shown in the figure) is implemented between the screens of the two spatial light modulators.}
\end{figure*}


{\em Device-independent dimension witness for dimension 6.---}To witness the generation of six-dimensional quantum systems and discriminate between classical and quantum systems of dimension 6, we implement the DI DW of the family $I_N$ \cite{GBHA10} that requires the smallest number of preparations and measurements and is capable to achieve this goal. This DI DW is $I_7$ and is defined within the following scenario: physical systems prepared in states $\rho_x$ with $x \in \{1,\ldots,N\}$ are submitted to a measurement $y \in \{1,\ldots,N-1\}$ with possible outcomes $b \in \{-1,1\}$; see Fig. \ref{setup} (a). The point is to chose these preparations and measurements in such a way that an external observer does not need to know which are these states and measurements to reach conclusions about the dimension of the prepared systems. For that, $I_N$, with $N=2,3,\ldots$, are specific linear combinations of conditional probabilities $P(b|x,y)$ of obtaining the outcome $b$ when measure $y$ is performed upon the state $\rho_x$, and have different upper bounds depending on the dimension of the prepared systems and on whether the systems are classical or quantum. Specifically, $I_N $ are of the form
\begin{equation}
 I_N \equiv \left|\sum^{N-1}_{j=1} E_{1j} + \sum^{N}_{i=2}\sum^{N+1-i}_{j=1} \alpha_{ij}E_{ij}\right|,
\end{equation}
where $E_{xy}\equiv P(+1|x,y)-P(-1|x,y)$ and $\alpha_{ij}=1$ if $i+j\leq N$ and $\alpha_{ij}=-1$, otherwise. For classical systems of dimension $d \leq N-1$, the maximum value of $I_N$ is $L_d=\frac{N(N-3)}{2}+2d-1$. $I_N$ is specially designed to certify systems of dimension $d=N-1$ and discriminate between classical and quantum systems of dimension $d=N-1$, since a quantum system of dimension $d=N-1$ can give a value of $I_N$ higher than $L_{d=N-1}$.

Here, we use $I_7$ to certify the generation of six-dimensional qudit systems. $I_7$ is given by
\begin{eqnarray}
 I_7 &\equiv& |E_{11} + E_{12} + E_{13} + E_{14} + E_{15} + E_{16} + E_{21} \nonumber \\
 && + E_{22} + E_{23} + E_{24} + E_{25} - E_{26} + E_{31} + E_{32} \nonumber \\
 && + E_{33} + E_{34} - E_{35} + E_{41} + E_{42} + E_{43} - E_{44} \nonumber \\
 && + E_{51} + E_{52} - E_{53} + E_{61} - E_{62} - E_{71}|.
\end{eqnarray}

Unlike for the classical limits, for the quantum limits of $I_7$ there are not known analytical expressions. Here we obtain these limits using the numerical technique of the conjugated gradient \cite{metGC}. We focus on real $d$-dimensional states, hence, characterized by $d-1$ parameters, and use the following parametrization based on $d$-dimensional spherical coordinates:
\begin{subequations}
\label{parametrization}
\begin{align}
 \lambda_1 &= \cos\phi_1, \\
 \lambda_j &= \cos\phi_j \prod^{j-1}_{k=1} \sin\phi_k, \\
 \lambda_{d-1} &= \cos\phi_{d-1} \prod^{d-2}_{k=1} \sin\phi_k, \\
 \lambda_{d} &= \sin\phi_{d-1} \prod^{d-2}_{k=1} \sin\phi_k,
\end{align}
\end{subequations}
with $j=2,\ldots,d-2$, and where the quantum states to be prepared are $\rho_x = |\Psi_x\rangle\langle\Psi_x|$, with
\begin{equation}
 |\Psi_x\rangle \equiv \sum_{i=0}^{d-1} \lambda_{i+1}^{(x)}|i\rangle.
\end{equation}
The measurements to be made are given by $M_y^+~=~|\Psi_y\rangle\langle\Psi_y|$, where $|\Psi_y\rangle = \sum_{i=0}^{d-1} \lambda_{i+1}^{(y)}|i\rangle$.

Applied to estimating bounds for $I_7$, the conjugated gradient method consists of calculating the local gradient in a point of the $I_7$ parameter space, defined by the angles $\phi_i$ in Eqs.~(\ref{parametrization}), in order to reach the closest maximum point of $I_7$. This maximum is reached when the gradient is null. To map all the local maxima and decide which is the global one, we ran the conjugated gradient algorithm for a large uniform sample of points in the parameter space ($5 \times 10^8$ trials) for each dimension considered. Table~\ref{Table1} shows the maximum values that $I_7$ can take, depending on the dimension of the prepared quantum states. From Table~\ref{Table1} one can see that, if the experimental value of $I_7$ is greater than $24.8987$, then it is guaranteed that the systems have at least dimension 6. To discriminate between classical systems of dimension 6 and quantum systems of dimension 6, one must observe an experimental value greater than $25$.


\begin{table}[t]
\centering
\begin{tabular}{ccc}
$d$ & $I_{7c}$ & $I_{7q}$ \\ \hline
2 & 17 & 17.3976 \\
3 & 19 & 20.7085 \\
4 & 21 & 23.2167 \\
5 & 23 & 24.8987 \\
6 & 25 & 26.1017 \\ \hline
\end{tabular}
\caption{Limits of $I_7$ for classical ($I_{7c}$) and quantum ($I_{7q}$) systems of dimension $d$.}
\label{Table1}
\end{table}


The states and measurements that lead to the maximum value of $I_7$ when the prepared systems are quantum systems of dimension $6$ are presented in Tables~\ref{Table2} and \ref{Table3}, respectively. Notice that, with the parametrization \eqref{parametrization}, the number of parameters that have to be experimentally controlled in order to test $I_N$ grows as $2N^2-5 N+2$; that is, to certify dimension $d$ using $I_N$, one needs to control $2 d^2-d-1$ parameters: $N(d-1)$ parameters related to the $N$ states to be prepared, plus $(N-1)(d-1)$ parameters related to the $N-1$ required projections.


\begin{table}[t]
\centering
\begin{tabular}{cccccc}
$x$& $\phi_{1}^{(x)}$ (rad) & $\phi_{2}^{(x)}$ (rad) & $\phi_{3}^{(x)}$ (rad) & $\phi_{4}^{(x)}$ (rad) & $\phi_{5}^{(x)}$ (rad) \\ \hline
1 & 4.8501 &  1.8679 & 5.1341 & 1.5056 &  4.4493 \\
2 & 1.7085 & -0.1307 & 2.9637 & 0.2325 &  1.0858 \\
3 & 1.4347 &  1.3779 & 5.0763 & 4.8358 &  6.2086 \\
4 & 4.5814 &  1.4557 & 0.6297 & 4.5888 & -0.1338 \\
5 & 4.8343 &  1.8268 & 4.9946 & 0.5796 &  1.6284 \\
6 & 4.6016 &  1.3527 & 4.2510 & 0.5860 &  4.9691 \\
7 & $\pi$  &  2.2358 & 5.2280 & 2.8465 &  0.5110 \\ \hline
\end{tabular}
\caption{Orientations of the states that maximize $I_7$ while considering six-dimensional qudit states.}
\label{Table2}
\end{table}


\begin{table}[tb]
\centering
\begin{tabular}{cccccc}
$y$& $\phi_{1}^{(y)}$ (rad) & $\phi_{2}^{(y)}$ (rad) & $\phi_{3}^{(y)}$ (rad) & $\phi_{4}^{(y)}$ (rad) & $\phi_{5}^{(y)}$ (rad)\\ \hline
1 & 0      & 6.0542 &  3.8912 & 1.8371 &  0.378 \\
2 & 1.3491 & 1.3527 &  2.0322 & 2.5556 &  1.8275 \\
3 & 1.7849 & 1.7926 &  1.3549 & 3.6187 &  1.6015 \\
4 & 1.3718 & 1.5194 & -0.5647 & 4.8676 &  3.2279 \\
5 & 1.3932 & 1.4595 &  5.0187 & 4.8270 & -0.1914 \\
6 & 1.7268 & 6.4857 &  5.7731 & 1.2669 &  4.4619 \\ \hline
\end{tabular}
\caption{Orientations of the measurements that maximize $I_7$ while considering six-dimensional qudit states.}
\label{Table3}
\end{table}


{\em Experimental setup.---}In our experiment, the prepared quantum states are encoded in the orbital angular momentum (OAM) of single photons \cite{Alle92}. This degree of freedom is related to the spatial modes of electromagnetic radiation and provides an infinite dimensional Hilbert space that allows us to implement qudits of, in principle, arbitrary $d$ using single photons. OAM is attracting great attention in the last decade for its applications in a variety of disciplines such as biophysics \cite{Padg11,Grie03,Fran08}, metrology \cite{Damb13nc}, astronomy \cite{Tamb11}, quantum communication \cite{Moli07,Damb12}, and fundamental quantum physics \cite{Naga12,Damb13}. Here we take advantage of the wide alphabet provided by OAM to encode six-dimensional single photon quantum states by choosing as a logical basis the following OAM eigenstates subset: $\{\ket{-3}_{O},\ket{-2}_{O},\ket{-1}_{O},\ket{1}_{O},\ket{2}_{O},\ket{3}_{O}\}$, where $\ket{l}_{O}$ identifies the state of a photon with $l\hbar$ of orbital angular momentum.

Figure~\ref{setup} (b) illustrates the experimental setup. Single photons in the fundamental TEM00 Gaussian state ($l=0$) are prepared in the desired OAM superposition state by means of a spatial light modulator (SLM1). This device modulates the phase wave front according to computer generated holograms specifically calculated to maximize the state fidelity \cite{qusix}. After the state generation, a second spatial light modulator (SLM2), is used in combination with a single mode fiber and single photon detector to perform a projective measurement on the desired state. The overall setup implements a DI DW, see Fig.~\ref{setup} (a), where each button on the state generation box corresponds to a different hologram to be displayed on SLM1 and each button on the measurement box corresponds to a different hologram to be displayed on SLM2. Using this setup we are able to generate and project over all the states needed to implement $I_7$. The generation and measurement processes are completely automated and computer controlled.

The states and measurements adopted to obtain the maximum value of $I_{7}$ are reported in Tables~\ref{Table2} and \ref{Table3}. For each one of the 6 projectors $M_y^+$ we need to find a basis which allows us to measure the probabilities $P(b|x,y)$. From the experimental point of view, the best choice is to complete the 6 bases by introducing the minimum number of extra projectors. Hence, we found 18 extra projectors that we used to complete the 6 bases.


{\em Experimental results.---}In order to check the quality of the experimental projectors, for each $M_y^+$ we measured the fidelity $F(M_y^+)=|\bra{\Psi_y}\Psi_y\rangle|^{2}$, i.e., the probability of obtaining a photon count when a state is projected over itself. An average fidelity of $F=(99.10 \pm 0.02)\%$ is obtained removing the contribution of the dark counts (DC).

As a second step, we measured all the probabilities needed for $I_7$, obtaining a value of
\begin{subequations}
\begin{align}
I_{7}&=25.95 \pm 0.02,\\
I_{7}^{\rm DC}&=25.44 \pm 0.02,
\end{align}
\end{subequations}
with and without removing dark counts, respectively. Errors are calculated by considering a Poissonian statistics for photon counts. The fact that the experimental value of $I_7$ is higher than 25 certifies the generation of six-dimensional quantum states; see Table~\ref{Table1}.

To show that classical states of dimension 6 actually produce smaller values for $I_7$, we prepared classical states defined as
\begin{equation}
 \rho_x=\sum_{i=0}^{d-1}\lambda_{i+1}^2\ket{i}\bra{i}.
\end{equation}
For maximizing $I_7$ with {\em classical} states we used the orientations obtained again by means of the conjugated gradient method. Like in the quantum case, in the classical case we measured fidelities for all the projectors involved in the DW. The averaged value is $F=(99.000 \pm 0.005)\%$, removing dark counts. The $I_7$ measured values are
\begin{subequations}
\begin{align}
 I_{7}&=24.825 \pm 0.004,\\
 I_{7}^{\rm DC}&=24.733 \pm 0.006,
\end{align}
\end{subequations}
with and without removing dark counts, respectively.

\begin{figure}[t]
 \centering
\includegraphics[scale=0.11]{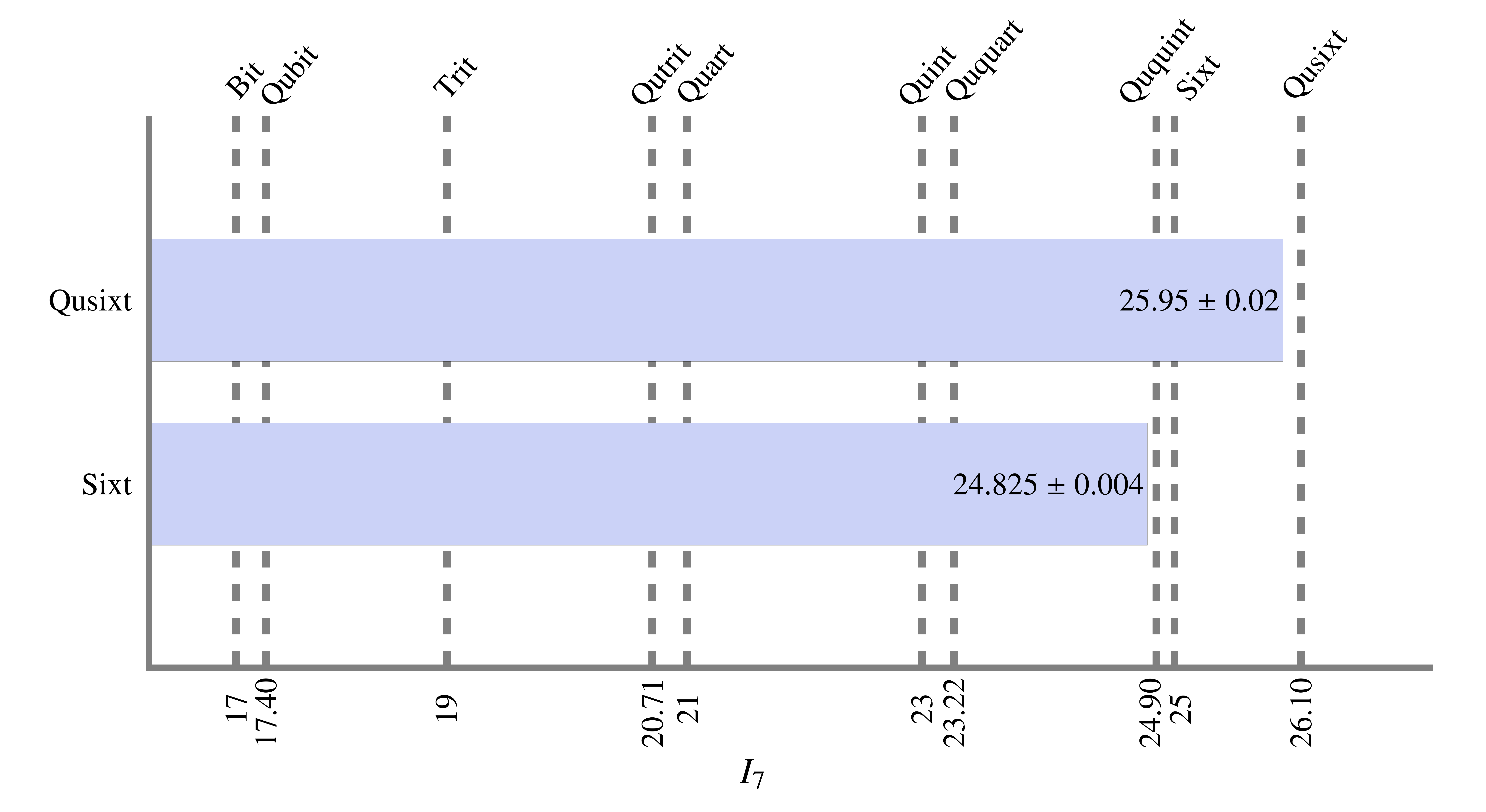}
 \caption{Experimental results for the test of the DI DW $I_7$ using quantum systems of dimension 6 (qusixt) and classical systems of dimension 6 (sixt).}
 \label{Fig2}
\end{figure}

The experimental results are compared with the theoretical bounds in Fig.~\ref{Fig2}. Specifically, Fig.~\ref{Fig2} shows the experimental results obtained for the DW $I_7$ when classical and quantum systems of dimension $d=6$ were generated. The results show that the values obtained using 6-dimensional quantum states cannot be simulated with classical states of the same dimension.


{\em Scalability.---}The experimental technique adopted in this work allowed us to generate and measure six-dimensional systems with high fidelity. It is then worth to investigate how far this technique can be scaled up. For this purpose, we have performed some numerical simulations in order to study the experimental requirements to discriminate between a quantum system of dimension $d$ and a quantum system of dimension $d-1$ using $I_N$. Specifically, we have studied the classical and quantum bounds of $I_N$ from $d=3$ up to $d=19$.

To do so, we considered the expected value of $I_{N}$ to lie in the range $[I_{N}^{\rm min},I_{N}^{\rm max}]$, where $I_{N}^{\rm min}= F I_{Nq}$ and $I_{N}^{\rm max}=F I_{Nq}+(1-F)(N+2)(N-1)/2$, where $I_{Nq}$ is the numerical quantum limit and $F$ is the mean fidelity over all the states. These bounds were obtained by considering that the lowest and the highest algebraic values of $I_{N}$, which are $0$ and $(N+2)(N-1)/2$, respectively, are obtained with probability $1-F$. Dimension witnessing is possible until $I_{N}^{\rm max}$ of a $d-1$ dimensional quantum system is lower than $I_{N}^{\rm min}$ of a $d$ dimensional one. Since the accuracy required for dimension $d=13$ is still compatible with our experimental errors, it is realistic to conjecture that our technique works for this dimension, assuming that $F$ remains constant. This assumption is approximately satisfied, at least up to $d=16$, in certain implementations \cite{q16}. A similar analysis shows that, even by considering a more adversarial scenario in which fidelity decreases to $F=98\%$, witnessing the dimension in a DI way would be still possible up to dimension $d=10$.

On the other hand, we have observed that, at least for the dimensions we have studied, for a given dimension $d$, the difference between $I_{d+1}$ for a quantum and a classical system increases with $d$. This makes the discrimination between the quantum and the classical cases easier for higher dimensions.


{\em Conclusions.---}High-dimensional quantum systems will be needed both for specific applications of quantum information processing \cite{CDNS11} and for increasing the transmission rate of quantum information, as required in realistic communication scenarios. Developing experimental methods to certify the generation of high-dimensional quantum systems, and especially of high-dimensional photonic qudits in which quantum information is encoded in the transverse spatial mode of light, is therefore of utmost importance. Moreover, for many applications, it is desirable to provide this certification in a device-independent manner, that is, without making assumptions about the internal functioning or the quality of the devices used to prepare and measure the systems. The question was whether device-independent certification of quantum systems of high dimension, and, in particular, of high-dimensional photonic qudits, was experimentally feasible. Here we have shown that it is. We have certified in a device-independent manner photonic qudits of dimension 6, by showing that the observed statistics cannot be produced either by quantum systems of lower dimensions or by classical systems of dimension 6. We have also shown that the same method can be used to certify even higher dimensions in a device-independent manner. Specifically, we have shown that, encoding high-dimensional quantum information in the transverse spatial mode of light, the accuracy needed to assess dimensions up to 13 (10 in a more adversarial scenario) is still within the experimental error of our experiment. These results demonstrate the feasibility of the device-independent approach for realistic high-dimensional quantum information processing.


\begin{acknowledgments}
The authors thank Antonio Ac\'{\i}n for useful comments. This work was supported by Project No.\ FIS2011-29400 (MINECO, Spain) with FEDER funds, the FQXi large grant project ``The Nature of Information in Sequential Quantum Measurements,'' the Brazilian program Science without Borders, FIRB Futuro in Ricerca-HYTEQ, the Chillean Grants FONDECYT 1120067, Milenio~P10-030-F, and PIA-CONICYT PFB0824, the ERC Advanced Grant QOLAPS and Starting Grant 3D-QUEST (3D-Quantum Integrated Optical Simulation; Grant Agreement No. 307783). J.F.B. acknowledges the financial support of CONICYT.
\end{acknowledgments}



\end{document}